\documentclass[aps,prb,reprint,10pt,fleqn,showpacs,showkeys,floatfix,superscriptaddress]{revtex4-1}
\usepackage{graphicx,color}
\usepackage{amssymb}
\usepackage{amsmath}
\usepackage{bm}
\usepackage{dcolumn}
\usepackage[normalem]{ulem}

\def\iTone{T_1^{-1}}

\begin{document}

\title{Antiferromagnetic Ordering in MnF(salen)}

\author{Erik \v{C}i\v{z}m\'{a}r}
\affiliation{Department of Condensed Matter Physics, P.~J.~\v{S}af\'{a}rik University,
Park Angelinum 9, 041 54 Ko\v{s}ice, Slovakia}
\author{Olivia N. Risset}
\affiliation{Department of Chemistry, University of Florida, Gainesville, FL 32611-7200, USA}
\author{Tong Wang}
\affiliation{Department of Physics and Astronomy, University of California, Los Angeles, CA 90095-1547, USA}
\author{Martin Botko}
\affiliation{Department of Condensed Matter Physics, P.~J.~\v{S}af\'{a}rik University,
Park Angelinum 9, 041 54 Ko\v{s}ice, Slovakia}
\author{Akhil R. Ahir}
\affiliation{Department of Chemistry, University of Florida, Gainesville, FL 32611-7200, USA}
\author{Matthew J. Andrus}
\affiliation{Department of Chemistry, University of Florida, Gainesville, FL 32611-7200, USA}
\author{Ju-Hyun Park}
\affiliation{National High Magnetic Field Laboratory, Florida State University, Tallahassee, FL 32310-3706, USA}
\author{Khalil A. Abboud}
\affiliation{Department of Chemistry, University of Florida, Gainesville, FL 32611-7200, USA}
\author{Daniel R. Talham}
\affiliation{Department of Chemistry, University of Florida, Gainesville, FL 32611-7200, USA}
\author{Mark W. Meisel}
\affiliation{Department of Physics and National High Magnetic Field Laboratory, University of Florida, Gainesville,
FL 32611-8440, USA}
\author{Stuart E. Brown}
\affiliation{Department of Physics and Astronomy, University of California, Los Angeles,
CA 90095-1547, USA}

\date{\today}

\begin{abstract}
Antiferromagnetic order at $T_{\mathrm{N}} = 23$~K has been identified in Mn(III)F(salen),
salen = H$_{14}$C$_{16}$N$_2$O$_2$, an $S = 2$ linear-chain system.  Using single crystals,
specific heat studies performed in magnetic fields up to 9~T revealed the presence of a
field-independent cusp at the same temperature where $^1$H NMR studies conducted at 42~MHz
observed dramatic changes in the spin-lattice relaxation time, $T_1$, and in the linewidths.
Neutron powder diffraction performed on a randomly-oriented, as-grown, deuterated (12 of 14 H replaced by d)
sample of 2.2~g at 10~K and 100~K did not resolve the magnetic ordering, while
low-field (less than 0.1~T) magnetic susceptibility studies of single crystals and 
randomly-arranged microcrystalline samples reveal subtle features associated with the transition.
Ensemble these data suggest a magnetic signature previously detected at 3.8~T
for temperatures below nominally 500~mK is a spin-flop field of small net moments arising from
alternating subsets of three Mn spins along the chains.
\end{abstract}

\pacs{75.50.Ee, 75.40.Cx, 76.60.Es, 75.25.-j}

\maketitle

\section{INTRODUCTION}
After Haldane identified significant differences in the magnetic behavior of integer and half-interger,
Heisenberg, antiferromagnetic spins in one-dimension,\cite{Haldane1,Haldane2}
some time elapsed before Ni(C$_2$H$_8$N$_2$)$_2$NO$_2$(ClO$_4$), commonly referred to as NENP,
emerged as a model $S=1$ system\cite{Renard} that exhibiting no evidence of long-range
ordering down to at least 4~mK.\cite{Avenel1,Avenel2}  With a wide-range of work
reported on $S=1$ Haldane systems,\cite{Yamashita} the challenge of finding an $S=2$ Haldane system was
reportedly resolved with the identification of MnCl$_3$(bpy), bpy = C$_{10}$H$_8$N$_2$
(2,2$^{\prime}$-bipyridine),\cite{Goodwin-Sylva,Perlepes} as a Haldane gapped
system with nearest-neighbor interaction $J \approx 35$~K and no long-range order down to
30~mK.\cite{Granroth1}
However, high-field magnetization\cite{Hagiwara1} and EPR studies\cite{Hagiwara2} of as-grown
microcrystalline samples at low temperature, $T = 1.3$~K, provided evidence of a spin-flop transition
and the presence of antiferromagnetic resonance (AFMR) modes.  Recently with the use of
single crystals, long-range antiferromagnetic ordering has been identified in MnCl$_3$(bpy) near
11~K.\cite{Hagiwara3,Shinozaki} 

Although other candidate $S=2$ linear-chain materials have been
identified,\cite{Yamashita,GranrothPhD,Matsuhita,Leone,CizmarJMMM,Stock,Stone}
these systems possess long-range ordering, and evidence of a
gapped quantum spin liquid state just above the ordering was not detected in CrCl$_2$.\cite{Stone}
In parallel with these experimental studies,
theoretical activity to extend and explore the quantum spin properties
of antiferromagnetic $S=2$ spins in one-dimension is topical and
intense.\cite{PhysRevB.83.224417,
doi:10.1143/JPSJ.80.043001,PhysRevB.84.140407,PhysRevB.85.075125,PhysRevB.86.024403,
PhysRevB.87.045115,PhysRevB.87.235106,PhysRevB.91.205118}

Recently, a new $S=2$ linear-chain system, MnF(salen), salen = H$_{14}$C$_{16}$N$_2$O$_2$,
was synthesized, and the low-field, high-temperature magnetic properties were fit with
$J/k_B \approx 46$~K, while no evidence of a magnetic transition was detected down to 1.8~K.\cite{Birk}
This report motivated a suite of studies, including torque magnetometry and EPR on single crystals and
neutron scattering on a partially deuterated, as-grown, microcystalline powder-like sample,
and a brief report of the resulting data sets are published elsewhere.\cite{Park}
During the course of this work, specific heat and $^1$H NMR investigations were initiated,
and both experiments provided unambiguous evidence of long-range antiferromagnetic order at $T_{\mathrm{N}}=23$~K.
The purpose of this paper is to present these data sets and the corresponding analyses, which, when combined
with neutron powder diffraction at 10~K and 100~K and additional systematic studies of the low-field magnetic
response, provide insight into the nature of the magnetic transition.

\section{EXPERIMENTAL DETAILS and Results}

\subsection{Material Preparation}
All non-deuterated chemical reagents were purchased from Sigma-Aldrich, Alfa Aesar, or Tokyo Chemical Industry
(TCI) Company and used without further purification. The deuterated reagents were purchased from C/D/N Isotopes.
Using the facilities at the University of Florida, the hydrogenated samples of MnF(salen) were synthesized as
described elsewhere.\cite{Birk}  The identity and purity of the compound were confirmed by
X-ray diffraction and FT-IR spectroscopy. The deuterated sample MnF(salenH2-d12) was prepared following the
same procedure\cite{Birk} using deuterated precursors salicylaldehyde-d4
and 2,2-bis(salicylaldehyde)ethylenediimine acid-d12 (salenH2-d12).

\paragraph*{Synthesis of salicylaldehyde-d4.}
Salicylaldehyde-d4 was synthesized using a modification of the method reported elsewhere.\cite{Hofslokken}
Paraformaldehyde (110~mmol) was added to phenol-d6 (16~mmol), anhydrous magnesium dichloride (24~mmol) and
triethylamine (61~mmol) in 40~mL dry acetonitrile yielding a white suspension. The mixture was refluxed for 8~hours
and the resulting yellow suspension was allowed to cool overnight. Water (50~mL) was added and the mixture was
acidified with aq.~HCl to reach pH = 2. Water was further added to increase the mixture volume to 150~mL.
The product was extracted with ether ($3\times200$~mL), dried with magnesium sulfate and evaporated.
The resulting orange oil was purified by flash chromatography on silica gel to give a light yellow oil.
Yield = 8.9 mmol (56\%). $^1$H NMR (300 MHz, CDCl$_3$): δ 9.83 (s, $^1$H, CHO), 10.95 (s, $^1$H, OH).
FT-IR (cm$^{-1}$): 3442(broad), 2970(w), 2858(w), 1647(s), 1602(m), 1558(s), 1390(s), 1289(s), 1189(m),
1125(s), 1055(m), 1032(m), 888(m), 857(w), 832(m), 816(m), 751(m), 721(s).

\paragraph*{Synthesis of 2,2-bis(salicylaldehyde)ethylenediimine acid-d12  (salenH2-d12).}
SalenH2-d12 was prepared using a conventional procedure for salen synthesis based on the condensation of
salicylaldehyde with ethylenediamine.  To a boiling ethanol solution (20~mL) containing salicylaldehyde-d4 (14~mmol),
ethylene-d4-diamine (7~mmol) was added dropwise under vigorous stirring. After 10~min, the mixture was cooled
in an ice/water bath and the product was filtered, washed with cold ethanol, and air-dried yielding bright
yellow flaky crystals. Yield: 5.6 mmol (80\%). $^1$H NMR (300 MHz, CDCl$_3$): δ 8.58 (s, $^2$H, =CH),
13.37 (s, $^2$H, OH).  FT-IR (cm$^{-1}$): 3421(broad), 1631(s), 1591(s), 1551(s), 1444(s), 1395(s), 1333(s),
1300(s), 1220(m), 1151(m), 1098(s), 1052(s), 1025(m), 975(m), 857(s), 809(m), 767(m), 737(m), 588(s).

\subsection{Specific heat}

\begin{figure}[!]
\includegraphics[width=\linewidth]{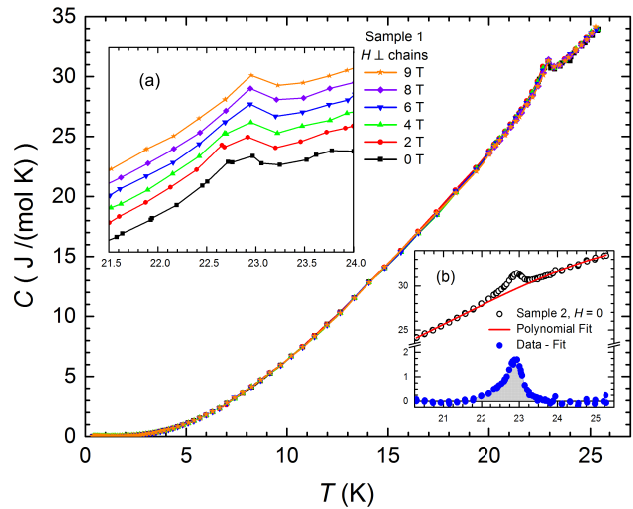}
\caption{\label{fig:Cv}  (Color online)  The temperature dependence of the
addenda-corrected specific heat measured in magnetic fields $(0\leq \mu_{\mathrm{o}}H \leq 9$~T)
with $H$ oriented nominally perpendicular to the chains is shown for Sample 1 (0.8~mg).
Inset (a) shows an expanded view of the the data in the vicinity
of the transition, and these data have been vertically shifted for clarity.  The solid
lines in the main panel and (a) are straight connections between data points.  Inset (b)
shows the data for Sample 2 (0.67~mg) in zero applied field.  A polynomial fit to the data away from
the transition region is shown by the solid (red) line.  The results of subtracting the
fit line from the data are shown at the bottom of the plot.  The shaded region provides 
(\emph{de facto} the area under a $C/T$ curve) an
estimate of the entropy change in the vicinity of the transition of 48 mJ/(K mol) or
0.4\% of the total magnetic entropy of $R\ln(5) = 13.38$~J/(K mol).}
\end{figure}

The specific heat studies used the facilities located at P.~J.~\v{S}af\'{a}rik University and were made with
several single crystal samples. In the temperature range between 0.38~K and 40~K,
a commercial system (Quantum Design PPMS) equipped with a $^3$He insert was used while employing the
relaxation method technique. Two independent measurements were made, where the first used one single crystal
of mass 0.8~mg and the second one employed two crystals of total mass 0.67~mg, and these data are
shown in Fig.~\ref{fig:Cv}.  Below 1~K, the experiment was conducted in a $^3$He-$^4$He dilution refrigerator 
(Air Liquide Minidil) using a 2.1~mg single crystal.
The sample was attached using Apiezon N vacuum grease to the home-made calorimeter consisting of a
sapphire substrate, which also supported a thin-layer RuO$_2$ thermometer and a resistive heater.
The RuO$_2$ thermometer was calibrated against a commercial (Scientific Instruments (SI)) RuO$_2$-based
thermometer. The magnetoresistance changes of the SI thermometer were included using a known
correction from the literature.\cite{Watanabe}  The specific heat was measured down to 60~mK by standard
dual-slope and heat-pulse methods,\cite{Hwang} which are similar to the ones used in PPMS instrument.

\subsection{NMR}
\def\iTone{$T_1^{-1}$}
\def\oneH{$^1$H}

The \oneH\ NMR studies were performed with standard four-phase cycling Fourier-transform spin-echo
techniques using a spectrometer and probe built at UCLA. The sample holder was constructed from
teflon and brass, and the coil was wound from teflon-coated wire.  The choice of these materials 
effectively reduced extraneous proton signals to an undetectable level. The circuit was bottom-tuned 
using fixed matching
and tuning elements, and the magnetic field set to the \oneH\ resonance condition (989.4~mT at 42.130~MHz)
using an electromagnet. The sample was cooled in a variable-temperature insert placed in a bucket
dewar with its tail between the magnet pole-faces. The sample studied was a single crystal of mass 5.1~mg,
with the field aligned orthogonal to the chain axis.

\begin{figure}[!]
\begin{center}
\includegraphics[width=8.6cm]{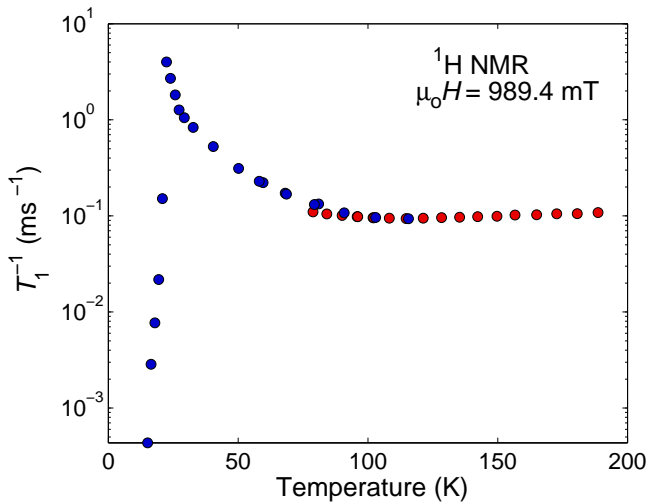}
\end{center}
\caption{The temperature dependence of the spin-lattice relaxation rate, \iTone, in an applied field of
989.4~mT and excitation frequency 42.13~MHz is shown for a single crystal with $H$ nominally aligned along
the chain direction when the cryogen employed was helium (blue) or nitrogen (red). 
The sharp drop below $T_{\mathrm{N}}=23$~K identifies the antiferromagnetic phase transition.}\label{fig:MnFiTone}
\end{figure}

The temperature dependence of the spin-lattice relaxation rate, \iTone, is shown in Fig.~\ref{fig:MnFiTone}.
These data exhibit a relatively weak variation with \iTone~$\sim 100$~s$^{-1}$ down to about 80~K. Further 
cooling leads to an monotonic increase of almost 2~orders of magnitude before precipitously dropping by 
many orders of magnitude below $T=23$~K. The increase over the range 23~K~$<T\simeq$~80~K signals the onset 
of antiferromagnetic correlations and the associated slowing of the fluctuating field. The drop in 
\iTone~below 23~K is accompanied by a loss of signal intensity associated with the discontinuous onset of 
line-broadening, most of which is outside the spectrometer operating bandwidth of order 100~kHz. 
The data were collected while cooling and warming and no hysteresis was detected.

\begin{figure}[!]
\begin{center}
\includegraphics[width=\linewidth]{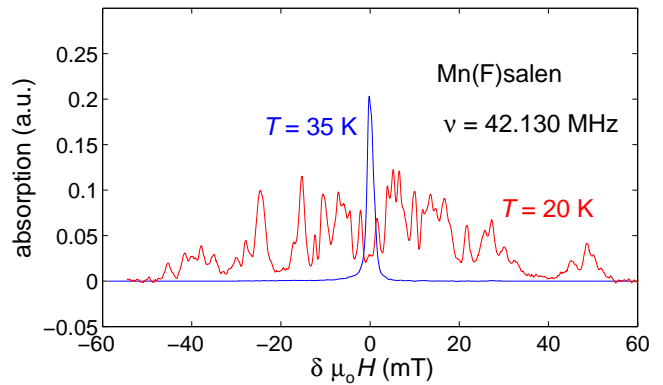}
\end{center}
\caption{The two spectra shown are from temperatures greater than (paramagnetic state, narrow line)
and less than $T_{\mathrm{N}}=23$~K. The low temperature spectrum is constructed from a spectral sum, where the
echo transients are recorded at 5~mT intervals. $\delta \mu_{\mathrm{o}}H = 0$ corresponds to an applied field 989.4~mT.}
\label{fig:MnFspectra}
\end{figure}

In Fig.~\ref{fig:MnFspectra}, two spectra, characteristic of temperatures above and below the
ordering temperature, are shown. For $T>T_{\mathrm{N}}$, the linewidth was approximately 50~kHz and exhibited unresolved
features that presumably result from inequivalent hyperfine couplings and internuclear spin-spin couplings.
For $T<T_{\mathrm{N}}$, the spectrum is much too broad for the pulse conditions ($p_1(\pi/2)=1.1$~$\mu$s, refocusing pulse
$p_2=0.7$~$\mu$s).  Thus, the full spectrum for the low-temperature phase was constructed from a sum of
field-swept spectra recorded at 5~mT intervals. Since the proton gyromagnetic ratio is $\gamma = 42.577$~MHz/T,
the equivalent steps in frequency correspond to slightly larger than 200~kHz, and therefore some spectral
distortions are certainly present. Nevertheless, in the data, there is a center of symmetry
about the unshifted position. Furthermore, the broadening was observed to onset discontinuously, indicative of a
first-order transition, and the overall symmetry of the lineshape is consistent with commensurate
magnetic ordering. The overall scale of the broadening ($\pm$~2~MHz) is consistent with direct
dipolar electron-nuclear spin coupling, when assuming ordered moments of order 1-2~$\mu_{\mathrm{B}}$ and a closest
\oneH-Mn distance of 3~\AA.

Returning to the nuclear spin-lattice relaxation, a temperature-independent result is expected in 
the case $k_{\mathrm{B}}T \gg z\,J$, \cite{Moriya:1956-1,Moriya:1956-2} where the hyperfine-field fluctuations 
predominantly originate with electronic spin $T_2$ processes determined by the exchange interaction $J$ and $z$ 
is the number of nearest neighbors. The relaxation rate in this limit can be estimated as
\begin{equation}
T_{1\infty}^{-1}\;=\;\frac{1}{2}\;\gamma^2\;\frac{h^2}{\omega_e}\;\;\;\;\;,
\end{equation}
with $\omega_e$ a characteristic frequency for the spectral density, given by $\omega_e^2=J^2zS(S+1)/\hbar^2$. 
To within geometric factors of order unity, the effective mean-square field for electron-nuclear dipolar coupling is
\begin{equation}
h^2\;=\;\frac{[2\pi]^{1/2}}{3}\frac{(\hbar\gamma_e)^2}{\omega_e}S(S+1)\sum_i<r_i^{-6}>\;\;\;\;,
\end{equation}
where the sum is over the proton-Mn distances for the $i^{th}$ Mn site. 
If we take the characteristic distance as 4~\AA, 
the result is $T_{1\infty}^{-1}\sim10$~s$^{-1}$, which is about an order of magnitude smaller than observed. 
Presumably the discrepancy arises from a combination of properly estimating the geometric factors, 
the field orientation, and/or the sum.

\subsection{Neutron Powder Diffraction}

\begin{figure}[!]
\includegraphics[width=\linewidth]{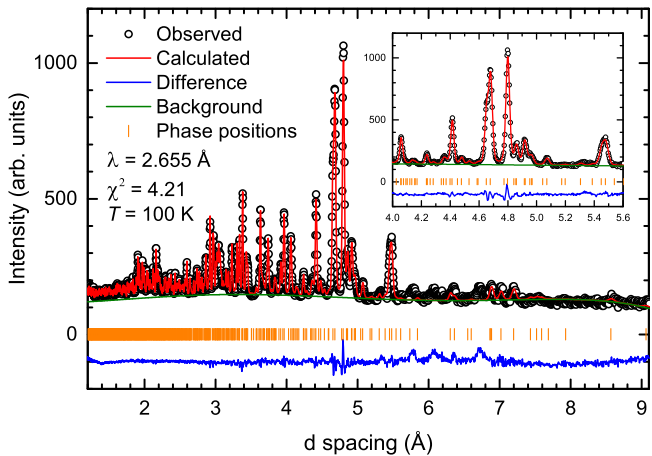}
\caption{\label{fig:NPD}  (Color online)  The intensity of the number of counts
as a function of the lattice spacing $d$ are shown for the NPD study performed 
on MnF(salenH2-d12) at 100~K. The inset shows an expanded view of the results.}
\end{figure}

Time-of-flight, powder neutron diffraction (TOF-NPD) studies were performed using the
POWGEN instrument at Spallation Neutron Source at Oak Ridge National Laboratory.
A deuterated MnF(salen) as-grown, randomly-arranged, powder-like sample with a mass of
2.2~g was placed in a vanadium can that was mounted in an ``Orange'' cryostat for collecting
data at 100~K, shown in Fig.~\ref{fig:NPD}, and 10~K. The diffraction patterns were
analyzed by Rietveld refinements using General Structure Analysis System package
(GSAS) and the graphical user interface (EXPGUI).\cite{GSAS,EXPGUI}
The data sets collected at both temperatures are essentially equivalent 
because, over the range of wave-vector measured, 
it was not possible to refine the 10~K structure with additional magnetic peaks.  
At 100~K, the unit cell parameters for MnF(salenH2-d12), Fig.~\ref{fig:NPD}, are 
$a = 10.01714$~\AA, $b = 15.43788$~\AA, $c = 16.00454$~\AA, $\alpha = 108.169^{\circ}$, 
$\beta = 103.936^{\circ}$, and $\gamma = 101.132^{\circ}$.  
The detailed analysis of both data sets will be given elsewhere.\cite{Akhil}

\subsection{Magnetometry}
\begin{figure}[!]
\includegraphics[width=\linewidth]{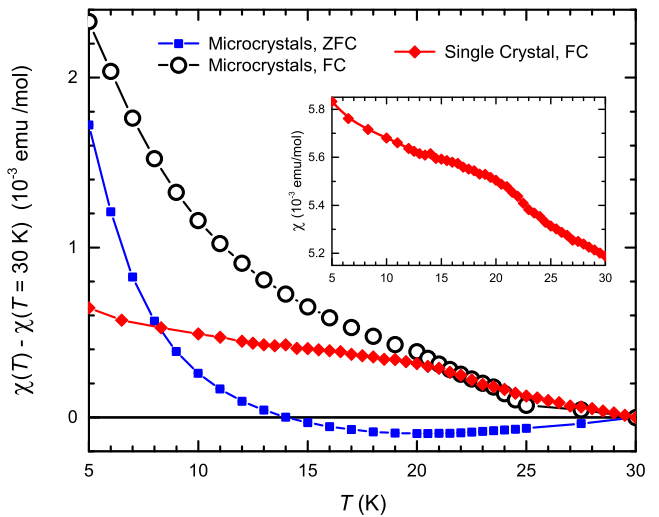}
\caption{\label{fig:Chi}  (Color online)  The magnetic susceptibility, referenced to the value
measured at 30 K, is shown as a function of temperature.  The red diamonds are the data, acquired
at 0.1~T with the field nominally along the chains, for the single crystal used in the NMR experiments,
and the inset shows the non-normalized values for 5~K $\leq T \leq 30$~K.  When zero-field-cooling (ZFC)
and field-cooling (FC) studies using 10~mT failed to reveal any detectable differences in the
microcrystalline sample as shown by the blue squares, the sample was FC in 7~T from room
temperature and subsequency measured while warming in 10~mT as marked by the open circles.  All of the
lines are guides for the eyes.}
\end{figure}

Two commercial magnetometers (Quantum Design MPMS-XL7), one at UCLA and the other at the
University of Florida, were used.  Previously, several batches, including the deuterated
one, of as-grown, randomly arranged microcrystals provided a response similar to the data
reported by Birk \emph{et al.}\cite{Birk}  Specifically, when using the Pad\'{e} approximations of
quantum Monte Carlo simulations,\cite{Law} the low-field magnetic susceptibility at high
temperatures was well fit when $J = 50 \pm 2$~K for $g = 2$.\cite{Park}  After the
discovery of the specific heat and NMR signatures of long-range antiferromagnetic
ordering at $T_{\mathrm{N}} = 23$~K, the magnetic susceptiblity studies were extended
to include a single crystal and zero-field-cooling (ZFC) versus field-cooling (FC) study of a
fresh (less than one-month old) as-grown microcrystalline sample.
The results shown in Fig.~\ref{fig:Chi} indicate the single crystal sample possesses
a subtle bump near 22~K when a field of 0.1~T is oriented nominally parallel to the chains.
Although ZFC and FC studies using 10~mT failed to reveal any variation of the magnetic
response for the microcrystalline sample, differences below 30~K were detectable when the
FC cycle was repeated in 7~T while cooling from room temperature and subsequently collecting
data while warming in 10~mT, Fig.~\ref{fig:Chi}.

\begin{figure}[h!]
\includegraphics[width=\linewidth]{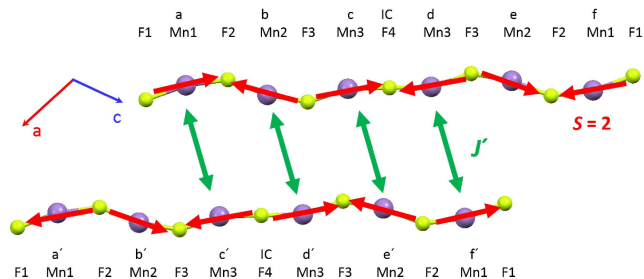}
\caption{\label{fig:spins}  (Color online)  A schematic of the Mn and F atoms of the nearest-neighbor
chains is shown, and these chains are not in the same plane as the image.  The a-axis and c-axis
projections are shown while the b-axis projection is hidden behind the c-axis vector.  The labels
of the Mn and F atoms follows the notation used by Birk \emph{et al.}\cite{Birk,Birk2} The suggested antiferromagnetic
arrangement shown by the red arrows is the same possibility described in the Supplemental Information, Fig.~SI2,
by Birk \emph{et al.}\cite{Birk}  The green arrows indicate the conjectured antiferromagnetic interchain coupling
$J^{\prime}$ due to the H$\cdot\cdot\cdot$H contacts of $< 3$~\AA~between interwoven salen molecules that are
not shown.  The proposed arrangement indicates a doubling of the lattice unit cell to generate the magnetic unit cell,
but the NPD study did not resolve the magnetic structure due to the limited $d$-spacing that was measured.
In addition, a net moment for every subset of three Mn atoms is antiferromagnetically arranged along the chains.
The torque magnetometry studies reported by Park \emph{et al.}\cite{Park} indicate a critical field of 3.8~T for
$T \lesssim 500$~mK, and this feature is the spin-flop field of the net moment of each subset of three Mn.}
\end{figure}

\section{Discussion and Summary}

Firstly, the specific heat data reveal the presence of a small bump at $T_{\mathrm{N}}=23$~K,
and this result is strikingly similar to the results reported for another $S=2$ quasi-linear chain material 
MnCl$_3$(bpy).\cite{Hagiwara3}  
In fact, the small value of the magnetic entropy removed at or below the ordering temperature is a signature of the 
low-dimensionality of the material since the majority of entropy is reduced by the low-dimensional 
short-range correlations at higher temperature.  Other examples of this situation have been reported for 
the one-dimensional antiferromagnet (CH$_3$)$_4$NMnCl$_3$,
commonly known at TMMC,\cite{Borsa} for the molecular magnet
[Fe$^{\mathrm{II}}$($\Delta$)Fe$^{\mathrm{II}}$($\Lambda$)(ox)$_2$(Phen)$_2$]$_n$,\cite{Ho} and the 
metal-organic framework Co$_4$(OH)$_2$(C$_{10}$H$_{16}$O$_4$)$_3$.\cite{Sibille}  
For MnF(salen), the magnetic contribution can not be separated from the phonon contribution,
so an accurate estimate of the total magnetic entropy removed below the ordering temperature is not possible.
The only theoretical work, which evaluates the contribution of three-dimensional ordering to the 
specific heat at the Ne\'el temperature, 
is reported for the two-dimensional square-lattice including interlayer interaction.\cite{Sengupta}
This theoretical work clearly shows how the entropy removed at the three-dimensional ordering transition 
is related to the strength of interlayer interaction, where weak-coupling between the planes generates a subtle anomaly
in the specific heat as only a tiny portion of entropy is removed. Conjecturing that this type of result might be
extended to one-dimensional chains, then one can infer the perturbation preventing the occurrence of a Haldane state for
$S=2$ chains might be very weak.  Additional theoretical consideration of this situation is warranted by the recent
findings reported here for MnF(salen) and elsewhere for MnCl$_3$(bpy).\cite{Hagiwara3}

Secondly, the NMR data from both single crystal and microcrystalline samples indicate a robust, first-order 
antiferromagnetic transition at $T_{\mathrm{N}}=23$~K.  The increase of \iTone~ by almost two orders of magnitude 
as the temperature is decreased toward $T_{\mathrm{N}}$ is consistent with the slowing of the moments due to 
increased correlations and the removal of entropy. These results indicate the importance of local probes, such as NMR and 
muon-spin rotation, that detect the formation of static moments and provide  transition signatures commonly missed 
by standard magnetometry techniques.  In addition, the difficulty of thermodynamically observing the transition 
of MnF(salen) is not caused by the competition between structural coherence and the magnetic correlation length.\cite{CizmarJMMM}

Finally, all of the results suggest MnF(salen) is antiferromagnetically ordered below 23~K, where 
the interchain coupling is antiferromagnetic, leading to the conjectured spin arrangement shown 
in Fig.~\ref{fig:spins}.  This arrangement was discussed by Birk \emph{et al.}\cite{Birk} and provides a 
potential reason why the NPD study of an unorientated microcrystalline sample did not unambiguously detect 
the magnetic spin arrangement.  Of course, an expanded $d$-spacing NPD study using a single-crystal sample 
could test this conjecture. 

\begin{acknowledgments}
This work was supported, in part, by the National Science Foundation via DMR-1105531 and DMR-1410343 (SEB),
DMR-1005581 and DMR-1405439 (DRT), DMR-1202033 (MWM), and DMR-1157490 (NHMFL), by the Slovak Agency for Research
and Development APVV-0132-11 and VEGA 0/0145/13 (EC), and by the Fulbright Commission Slovak Republic (MWM).  
Research at Oak Ridge National Laboratory’s
Spallation Neutron Source (SNS) was sponsored by the Scientific User Facilities Division,
Office of Basic Energy Sciences, U.S. Department of Energy.  The NPD experiments were performed by A.~Huq, who 
also provided significant analysis assistance.  Discussions with A.~Huq, S.~E.~Nagler, and the coauthors
of the earlier studies on this system\cite{Park} are gratefully acknowledged.
\end{acknowledgments}

\bibliography{MnFsalenv3}

\end{document}